\documentstyle[twoside,fleqn,espcrc2]{article}
\title{
\vspace{-2cm}
\begin{flushright}
KANAZAWA 96-10\\
July 1996
\end{flushright}
\vspace{.5cm}
Three topics of monopole dynamics in abelian projected QCD
\thanks{The authors of each part are 1. T.Suzuki, S.Kitahara, 
N.Nakamura, M.Sei and S.Kato, 2.  T.Suzuki, Y.Matsubara, N.Arasaki, S.Ejiri 
and S.Kitahara, 3. T.Suzuki and F.Shoji.}}
\author{Tsuneo Suzuki
\address{Department of Physics, Kanazawa University, Kanazawa 920-11, Japan}
, Yoshimi Matsubara
\address{Nanao Junior College, Nanao, Ishikawa 926, Japan},
 Shun-ichi Kitahara
\address{Jumonji University, Niiza, Saitama 352, Japan},
 Shinji Ejiri \hspace{-2mm}\addtocounter{address}{-3} 
\addressmark\hspace{2mm} 
,\\
 Naoki Nakamura \hspace{-2mm}\addtocounter{address}{-1} 
\addressmark\hspace{2mm}
, Fumiyoshi Shoji \hspace{-2mm}\addtocounter{address}{-1} 
\addressmark \hspace{2mm},
 Masafumi Sei \hspace{-2mm}\addtocounter{address}{-1} 
\addressmark \hspace{2mm},
Seikou Kato \hspace{-2mm}\addtocounter{address}{-1} 
\addressmark \hspace{2mm}
   and Natsuko Arasaki \hspace{-2mm}\addtocounter{address}{-1} 
\addressmark}

\begin{document}

\begin{abstract}
Three topics about monopole dynamics after abelian projection are 
reported. The first is the new and detailed analyses of $SU(2)$ monopole 
action obtained after the block-spin transformation on the dual lattice.
The $b=na(\beta)$ dependence for all couplings are well fitted with 
a universal curve. 
The distance dependence of the couplings is well reproduced by 
a massive propagator with the mass $m=0.8$ in unit of $b$.
The second is the $SU(3)$ monopole action recently obtained. The third is 
new interesting gauges showing abelian and monopole dominances as in 
the maximally abelian gauge.
\end{abstract}

\maketitle

\input epsf

\section{
Detailed analyses of $SU(2)$ monopole action
}

Confinement phenomena seem to 
be well reproduced by abelian link fields alone 
in the maximally abelian (MA) gauge 
in QCD\cite{kron,yotsu,hio,suzu93,suzu95}. 
The abelian dominance suggests 
the existence of an effective $U(1)$ theory 
$S_{eff}(u)$ 
describing confinement. 

After the abelian projection, one can separate out abelian link fields
$u(s,\mu)$ as  
$$ U'(s,\mu)= V(s)U(s,\mu)V^{\dagger}(s+\hat{\mu})
\equiv c(s,\mu)u(s,\mu).$$

Shiba and one of the authors (T.S.)\cite{shiba5} 
determined a monopole 
effective action defined as\cite{degrand} 
\begin{eqnarray*}
\exp(-S[k]) & = & \int Du \delta(k,u) \exp(-S_{eff}(u)),\\
\delta(k,u) & \equiv & \delta(
k_{\mu}(s) - \frac{1}{2}\epsilon_{\mu\nu\rho\sigma}
\partial_{\nu}m_{\rho\sigma}(s+\hat{\mu})).
\end{eqnarray*}
performing a dual transformation numerically. They also considered 
$n^3$ extended monopoles defined on a sublattice with the spacing 
$b=na$\cite{ivanenko}. 
This corresponds to making a block-spin transformation on 
the dual lattice as seen from 
\begin{eqnarray*}
e^{-S^{(n)}[k^{(n)}]}&=&(\prod_{s,\mu}\sum_{k_{\mu}(s)=-\infty}^{\infty})
    (\prod_s \delta_{\partial'_{\mu}k_{\mu}(s),0})\\ 
\times & &\prod_{s,\mu}\delta(k_{\mu}^{(n)}(s)-F(k))e^{-S[k]}, \\ 
\end{eqnarray*}
\vspace{-.5cm}
\begin{eqnarray*}
F(k)=
\sum_{i,j,l=0}^{n-1}k_{\mu}(ns+(n-1)\hat{\mu}+i\hat{\nu}
+j\hat{\rho}+l\hat{\sigma}).
\end{eqnarray*}
The monopole action adopted was composed of 12  two-point 
current-current 
interactions  $S[k] = \sum_i G_i S_i [k]$, the first of which is 
the self-coupling term $S_1[k]=\sum k_{\mu}^2(s)$.
The effective monopole actions $S^{(n)}[k^{(n)}]$ for $n=1\sim4$ 
were fixed from the emsemble $\{k^{(n)}_{\mu}(s)\}$ 
calculated  from vacuum 
configurations on $24^4$ lattice by extending the 
Swendsen method\cite{swendsen} for $\beta=2.5\sim 2.8$.
The monopole condensation\cite{thooft2}
 is shown to occur from energy-entropy balance
\cite{banks}.
However, the vacuum configurations for such $\beta$ have large
finite-size effects and the couplings $G_i$ ($i\ge 2$) have large errors.
 
It is the aim of this report to present new 
data for stronger coupling regions 
$\beta=2.0\sim 2.5$. Since the data are 
surprisingly clearer, we have adopted a monopole action with   
32 two-point couplings. The results are summarized as follows:
\begin{enumerate}
\item
The couplings up to the distance $\sqrt{8}$ are fixed 
clearly. 
\item
The distance dependence of the couplings is well reproduced by 
a massive propagator with the mass $m=0.8$ in unit of $b$.
The Smit-Sijs form (self-mass + Coulomb) is not good enough.
See Fig.\ \ref{dist}.
\begin{figure}[htb]
\epsfxsize=0.4\textwidth
 \begin{center}
 \leavevmode
  \epsfbox[0 133 595 729]{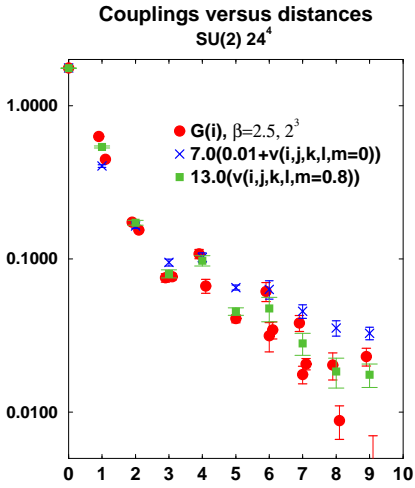}
 \end{center}
\vspace{-1.5cm}
 \caption{
Couplings versus squared distance between two currents.
}
\label{dist}
\end{figure}
\begin{figure}[tb]
\vspace{-1cm}
\epsfxsize=0.5\textwidth
 \begin{center}
 \leavevmode
  \epsfbox[0 133 595 729]{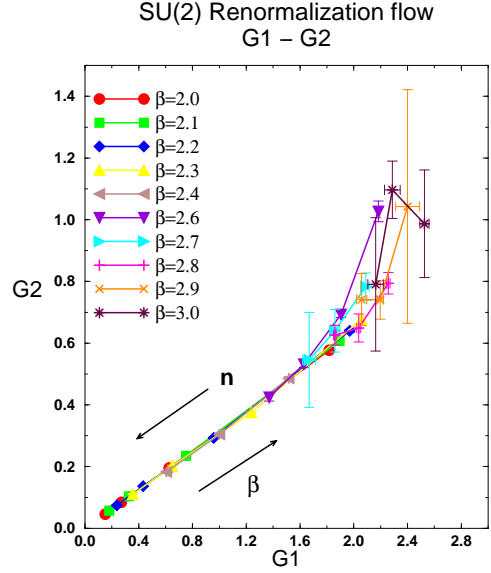}
 \end{center}
\vspace{-1.5cm}
 \caption{
The $G_1-G_2$ cross section of the renormalization flow of the block-spin 
transformation on the dual lattice.
}
\label{g12}
\end{figure}
\item
The scaling for fixed $b=na(\beta)$ 
 looks good for extendedness $n\ge 3$. 
\item
The $b$ dependence for all couplings are well fitted with 
a universal curve. See Fig\  \ref{g12} and also an example in 
Fig.\ \ref{bg11}.
\begin{figure}[htb]
\vspace{-1cm}
\epsfxsize=0.5\textwidth
 \begin{center}
 \leavevmode
  \epsfbox[0 133 595 729]{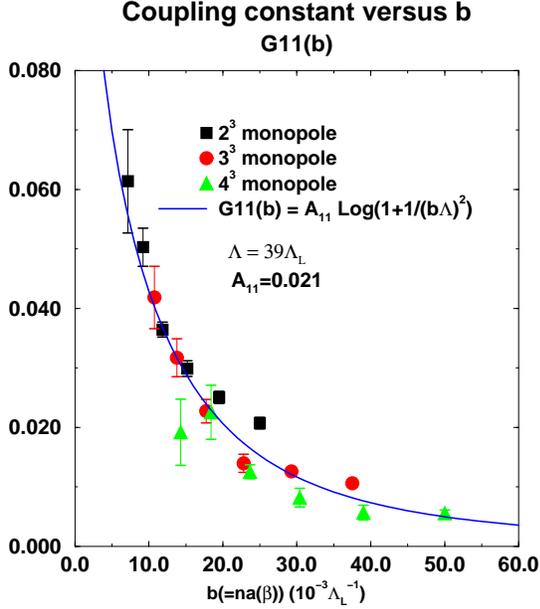}
 \end{center}
\vspace{-1.5cm}
 \caption{
The $b=na(\beta)$ dependence of the coupling $G_{11}$ in $SU(2)$.
}
\label{bg11}
\end{figure}
\item
The action is approximately fitted by the following form:
\begin{eqnarray*}
S(k)&=& g(b)\sum_{i}k_{\mu}(s)
\Delta^{-1}(\bar{i},m^2)k_{\mu}(s+\bar{i}),
\\
g(b)&=& C\log (1+1/(b\Lambda)^2),\\
m&\sim &0.8,\\
\Lambda&\sim &39\Lambda_L ,\\
C&\sim &4.9. 
\end{eqnarray*}
\end{enumerate}

\section{
The results of $SU(3)$ monopole action
}

In the case of $SU(3)$, 
there are two independent (three with one constraint 
$\sum_{i=1}^{3}k^i_{\mu}(s)=0$) currents.
When considering the two independent currents, their entropies are 
difficult to evaluate. Hence we first try to evaluate the effective 
monopole action, paying attention to only one monopole current.

The monopole action in $SU(3)$ QCD is obtained for
$\beta=5.0 \sim 6.0$\cite{arasaki,shiba5}. 
Lattice sizes considered are  $8^4\sim 24^4$ (for $T=0$ system).
The results are summarized as follows:
\begin{enumerate}
\item
The monopole actions for all extended monopoles are
 fixed in a compact form even in the 
scaling region. 
\item
Lattice-volume dependence is small.
\item
The total action is well approximated by the product of the self-coupling 
constant and the length $f_1\times L$.
\item
Energy-entropy shows monopole condensation as seen in Fig.\ \ref{f1be}.
\begin{figure}[htb]
\vspace{-1cm}
\epsfxsize=0.5\textwidth
 \begin{center}
 \leavevmode
  \epsfbox[20 133 615 729]{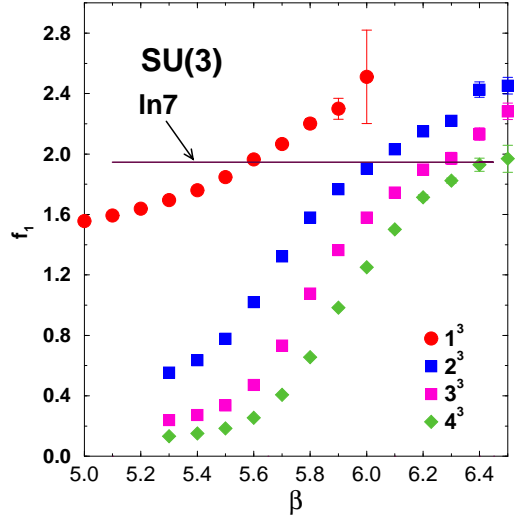}
 \end{center}
\vspace{-1.5cm}
\caption{
The $\beta$ dependence of the self-coupling term $f_1$ in $SU(3)$ case.
}
\label{f1be}
\vspace{-.5cm}
\end{figure}
\item
Scaling is not yet seen in $SU(3)$.
\item
There seems to be an infrared fixed-point at $f_i=0$ as in 
$SU(2)$.
\end{enumerate}

\section{
 New gauges showing abelian and monopole dominance
}

We have found two new gauges showing abelian and monopole dominances 
as in the maximally abelian (MA) gauge. The first one is 
the minimal abelian action (mAA) gauge
which minimizes the abelian action:
\begin{eqnarray*}
R&=&\sum_{s,\mu,\nu} \cos\theta_{\mu\nu}(s),\\
\theta_{\mu\nu}(s)&=&\theta_{\mu}(s)+\theta_{\nu}(s+\mu)
-\theta_{\mu}(s+\nu)-\theta_{\nu}(s).
\end{eqnarray*}
When we take 
 the $a\to 0$ continuum limit, we get  
\begin{eqnarray*}
\sum_{\mu,\nu}\partial_{\nu}f_{\mu\nu}(x)A^{\pm}_{\mu}(x)=0.
\end{eqnarray*}  

The second is the 
minimal abelian monopole density (mAMD) gauge which 
minimizes the abelian monopole density:
\begin{eqnarray*}
\rho&=&\sum_{s,\mu} |k_{\mu}(s)|.\\
k_{\mu}(s)&=&-\frac{1}{2}\epsilon_{\mu\nu\alpha\beta}\partial_{\nu}
\bar{\theta}_{\alpha\beta}(s+\mu) \\
\theta_{P}&=&\bar{\theta}_{P}+2\pi n_{P}
\end{eqnarray*}

We get the following:
\begin{enumerate}
\item
Abelian Wilson loops in mAA are enhanced as in MA, but they are smaller 
in mAMD.
\item
The string tension is reproduced by the abelian part and 
the monopole part as seen in Fig.\ \ref{string}..
\begin{figure}[htb]
\vspace{-1cm}
\epsfxsize=0.5\textwidth
 \begin{center}
 \leavevmode
  \epsfbox[20 133 615 729]{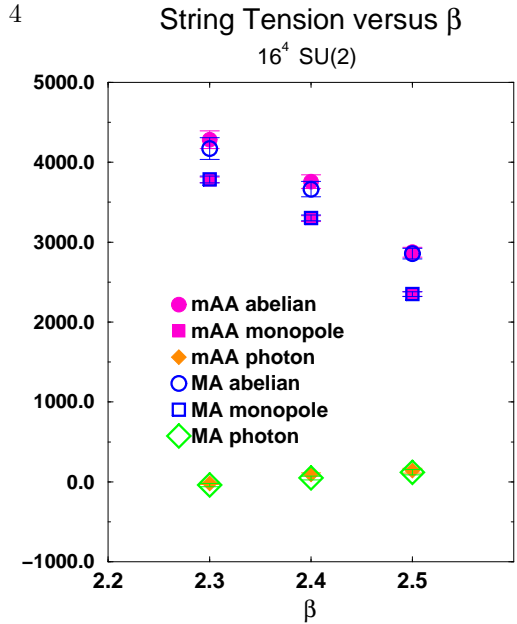}
 \end{center}
\vspace{-1.5cm}
\caption{
The string tension from the abelian , the monopole and the photon parts. 
}
\label{string}
\vspace{-.5cm}
\end{figure}
\item
The monopole densities in both new gauges are smaller than that in MA.
See Fig.\ \ref{density}.
\begin{figure}[htb]
\vspace*{-2.3cm}
\epsfxsize=0.5\textwidth
 \begin{center}
 \leavevmode
  \epsfbox[20 133 615 729]{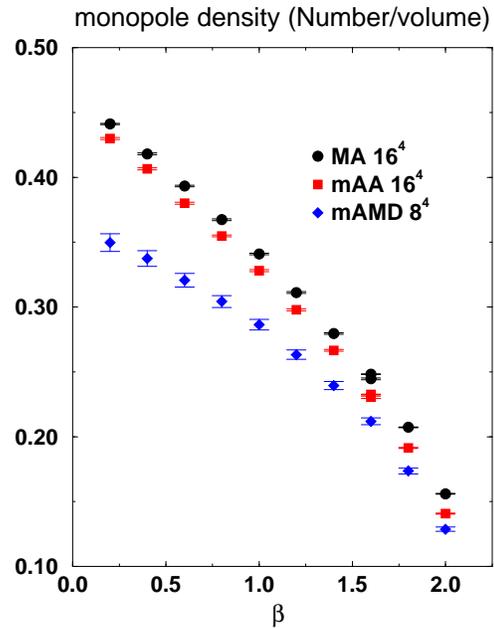}
 \end{center}
\vspace{-1.5cm}
\caption{
The monopole density in new gauges in comparison with that in MA.
}
\label{density}
\vspace{-.5cm}
\end{figure}
\item
The off-diagonal components in both gauges are less suppressed than in MA.
See for example Fig.\ \ref{hist}.
\begin{figure}[htb]
\epsfxsize=0.5\textwidth
 \begin{center}
 \leavevmode
  \epsfbox[20 133 615 729]{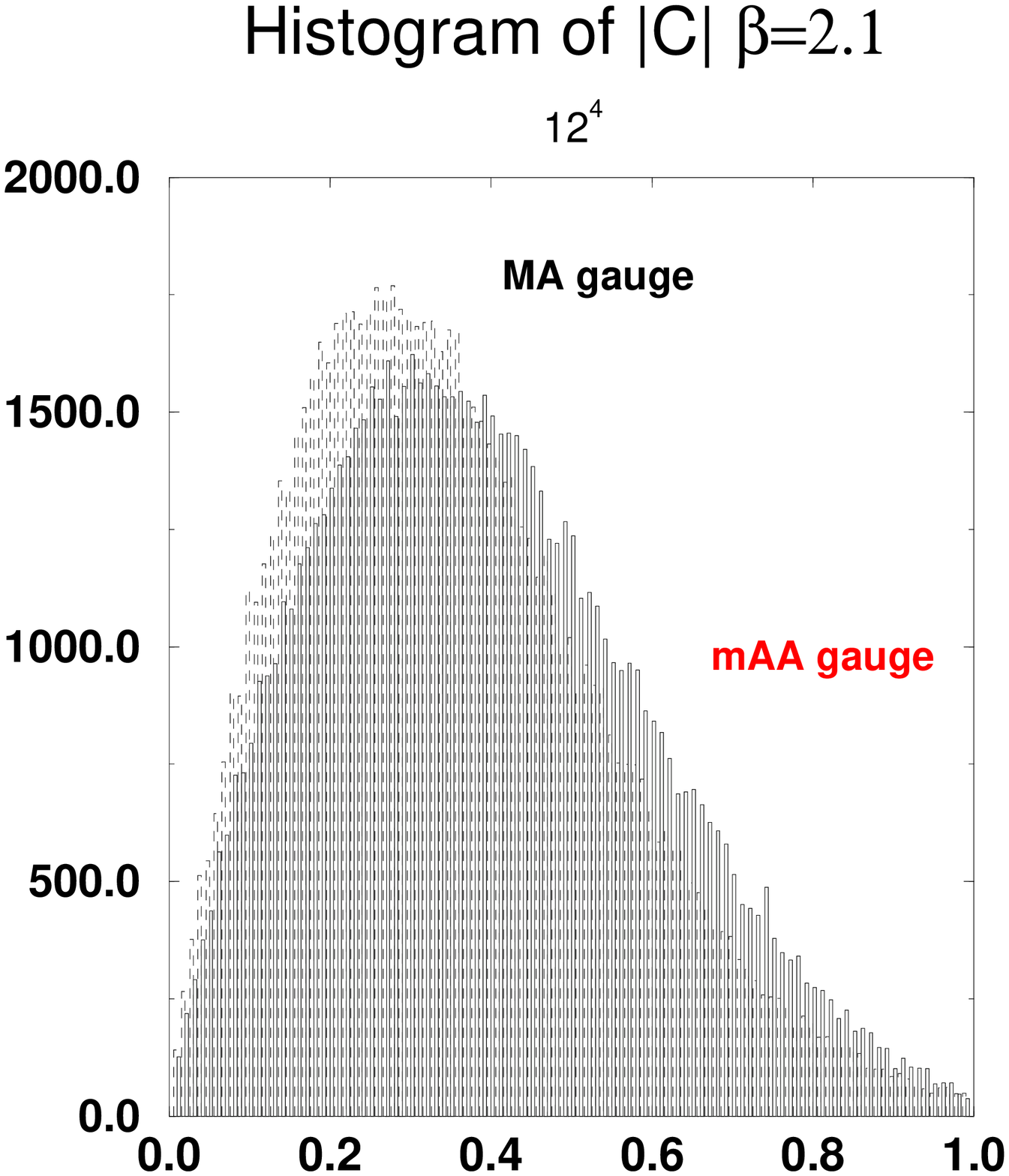}
 \end{center}
\vspace{-1.5cm}
\caption{
The histogram of the absolute value of the off-diagonal part in MA and mAA.
}
\label{hist}
\vspace{-.5cm}
\end{figure}
\item
Almost the same monopole actions are obtained in MA and mAA.
\end{enumerate}

The details of these three topics will be published elsewhere.

This work is financially supported by JSPS \\
Grant-in Aid for Scientific  
Research (B)\\ (No.06452028).

\end{document}